\documentclass[reprint,aps,pra,nofootinbib]{revtex4-2}

\usepackage{amsmath,amsfonts,amssymb}
\usepackage{mathtools}
\usepackage{bm}
\usepackage{tensor}
\usepackage{dcolumn}
\usepackage[colorlinks=true,allcolors=blue,linktocpage]{hyperref}
\usepackage[english]{babel}

\newcommand*{\differential}[1]{\mathinner{\mathrm{d}#1}}
\newcommand*{\ee}{\mathinner{\mathrm{e}}}

\newcommand{\Mnr}{M_\mathrm{nr}}

\newcommand{\pib}{\pi_\mathrm{bulk}}
\newcommand{\peff}{p_\mathrm{eff}}

\newcommand{\omegaeff}{\omega_\mathrm{eff}}
\newcommand{\omegaeffo}{\omega_{\mathrm{eff}, 0}}

\newcommand{\feq}{f_\mathrm{eq}}
\newcommand{\taueq}{\tau_\mathrm{eq}}
\newcommand{\Meq}{M_\mathrm{eq}}
\newcommand{\Ceq}{C_\mathrm{eq}}
\newcommand{\zeq}{z_\mathrm{eq}}
\newcommand{\Neq}{N_\mathrm{eq}}
\newcommand{\Teq}{T_\mathrm{eq}}

\begin{document}

\title{General relativistic non-ideal fluid equations for dark matter from a truncated cumulant expansion}

\author{Alaric~Erschfeld}
\email{erschfeld@thphys.uni-heidelberg.de}

\author{Stefan~Floerchinger}
\email{floerchinger@thphys.uni-heidelberg.de}

\author{Maximilian~Rupprecht}
\email{rupprecht@thphys.uni-heidelberg.de}

\affiliation{
Institut f\"{u}r Theoretische Physik, Ruprecht-Karls-Universit\"{a}t Heidelberg \\
Philosophenweg~16, D-69120 Heidelberg, Germany
}

\begin{abstract}
A new truncation scheme based on the cumulant expansion of the one-particle phase-space distribution function for dark matter particles is developed. Extending the method of moments in relativistic kinetic theory, we derive evolution equations which supplement the covariant conservation of the energy-momentum tensor and particle number current. Truncating the cumulant expansion we obtain a closed, covariant and hyperbolic system of equations which can be used to model the evolution of a general relativistic non-ideal fluid. As a working example we consider a Friedmann-Lema\^{i}tre-Robertson-Walker cosmology with dynamic pressure and solve for the time evolution of the effective equation of state parameter.
\end{abstract}

\maketitle

\section{Introduction}\label{sec:Introduction}
Observations indicate that a substantial part of the energy budget of the Universe is made up of dark matter. To a good approximation, it seems to be non-interacting except through gravitational interactions \cite{Kolb1990, Weinberg2008}. While there are various ideas as to the nature of dark matter, there is no verified microscopic theory so far. In the absence of a direct detection, its properties can only be constrained through cosmological and astrophysical observations. These are necessarily observations in the macroscopic domain, while one would ultimately be interested in a microscopic understanding in terms of a quantum field theory.
\par
In order to learn more about the properties of dark matter, one therefore turns to effective descriptions, such as fluid approximations or variants of kinetic theory. While isotropy and homogeneity constrain the cosmological background dynamics to be of perfect fluid type \cite{Peebles1980}, gravitational collapse is caused by fluctuations which are expected to generate non-vanishing shear stress at smaller scales and therefore require a non-ideal fluid description. For a consistent coupling to gravity through the Einstein field equations and in order to have a large scale description that obeys relativistic causality, one should aim for a relativistic fluid description. This is also important to understand the possible interplay with modified theories of gravity.
\par
Another striking example for the applicability of relativistic fluid dynamics are heavy-ion collisions, such as those studied at the Relativistic Heavy Ion Collider (RHIC) and the Large Hadron Collider (LHC). At macroscopic scales these exhibit collective fluid-like behaviour, such that the bulk dynamics can be described in terms of relativistic non-ideal fluid dynamics. This should be understood as a low-energy effective description of quantum chromodynamics in a out-of-equilibrium situation of high energy density \cite{Schaefer2009, Heinz2013, Gale2013, DeSouza2016}.
\par
First attempts to construct a theory of relativistic fluid dynamics were made by Eckart~\cite{Eckart1940} and later by Landau and Lifschitz~\cite{Landau1959}. However, these are known to be unstable and acausal \cite{Pichon1965, Hiscock1983, Hiscock1985} and can even have modes which propagate faster than light \cite{Hiscock1987}. In general, many so-called \textit{first-order theories} suffer from the fact that the equations of motion are parabolic in structure making them acausal \cite{Muronga2004}. Only recently, a class of first-order theories has been proposed that does not suffer from these problems \cite{Bemfica2018, Kovtun2019, Bemfica2019Nonlinear, Hoult2020}. While the energy-momentum tensor is expanded to first order in gradients around the ideal fluid form, this is done in such a way that the resulting equations of motion are second-order hyperbolic differential equations and can therefore obey relativistic causality.
\par
As another possibility to remedy these issues, extended theories have been proposed. In general, hydrodynamical descriptions are possible due to conservation laws. In the relativistic context, this is the covariant conservation of the energy-momentum tensor, sometimes supplemented by a covariantly conserved particle number current. In such a framework one has ten degrees of freedom in the components of the energy-momentum tensor and four in the particle number current, but only five equations of motion from energy-momentum and particle number conservation. Assuming that all the degrees of freedom are independent and dynamical, the system of equations is not closed and additional information is needed.
\par
The standard theory of hydrodynamics~\cite{Landau1959} provides such additional relations from an expansion of the conserved currents in terms of gradients of the fields that govern an ideal fluid, namely the fluid velocity and thermodynamic variables such an temperature and chemical potential. At a given finite order, this leads to additional constraint equations that related the components of the conserved currents to the hydrodynamic fields. Such a gradient expansion is particularly well motivated in the vicinity of thermal equilibrium or when interaction effects are so strong that they quickly drive the system back towards local equilibrium when the latter is violated as a consequence of fluid motion.
\par
In the context of dark matter, this close-to-equilibrium gradient expansion is not particularly well motivated, especially at late times and at small scales. Indeed, when interactions besides gravity are absent, one cannot assume that local thermal equilibrium is restored quickly. However, additional information to close the system of equations can also be provided in the form of additional evolution equations. This leads to a formalism with more dynamical variables or fluid fields, for example shear stress, bulk viscous pressure, heat current or particle diffusion current.
\par
Historically, such additional evolution equations have been first obtained from a kinetic theory approach where the system is characterised by a phase-space distribution function which obeys the relativistic Boltzmann equation. With the method of moments \cite{Grad1949} one can derive additional equations of motion in order to evolve the dynamical degrees of freedom and close the system by a systematic truncation, such as a derivative expansion. This has been done, most notably in the non-relativistic case by M\"{u}ller~\cite{Mueller1967} and in the relativistic case by Israel and Stewart~\cite{Isreal1976,Isreal1979Transient}, keeping terms up to second order in gradients, so-called \textit{second-order theories}. More modern approaches include derivative expansions, including all terms of second order in gradients \cite{Baier2008} and formulations derived from the relativistic Boltzmann equation, keeping all terms in the moment expansion \cite{Denicol2012DerivationOfTransient}. Recent applications of such approaches include references~\cite{Denicol2012DerivationOfFluid, Denicol2018, Denicol2019, Kurkela2018, Kurkela2019, Blaizot2018, Blaizot2020}.
\par
In the cosmological context, relativistic fluid dynamics have been extensively studied. These investigations range from ideal fluid dynamics \cite{Gallagher2018, Castiblanco2019}, to second-order theories \cite{Maartens1995, Zimdahl1996, Bemfica2019Causality} and to the Einstein-Vlasov system of equations \cite{Andreasson2011, Piattella2013, Piattella2016}. Further there have been investigations of effective fluid theories \cite{Baumann2012, Carrasco2012, Porto2014} and more recently a statistical field theory which maps to a non-equilibrium kinetic theory has been developed \cite{Viermann2015, Bartelmann2016, Bartelmann2019}.
\par
In the work presented here, we essentially adapt the method of moments and study a truncation of the \textit{cumulant expansion} of the phase-space distribution function for dark matter particles without collisions. In particular, we concentrate on a description of dark matter as classical particles. A more direct relation of the fluid picture to a quantum field theoretic description is left for future work (see also references~\cite{Friedrich2017, Friedrich2018}).
\par
In section \ref{sec:RelativisticKineticTheory} we present the relativistic Vlasov equation, which governs the evolution of the one-particle phase-space distribution function. We define moments and cumulants and study their behaviour under the Vlasov equation. Here, the first and second moment are the particle number current and energy-momentum tensor, respectively. We propose a truncation scheme in terms of the cumulant expansion of the distribution function and explicitly perform this truncation after the first and second cumulant. For these one can reconstruct a modified version of the phase-space distribution function. We discuss the non-consistency of these truncations, in the sense that they are not preserved by the Vlasov equation.
\par
In section \ref{sec:ClosedSystemOfEq} we derive closed equations of motion for the dynamical fluid fields parametrising the 14 independent degrees of freedom. To this end we use the covariant conservation of the first three moments and close the system of equations by neglecting the third cumulant. We discuss the hyperbolic structure of the system of equations, allowing for a causal description provided the characteristic propagation speeds are finite. By this we construct a closed, covariant and hyperbolic system of partial differential equations which can be used to model a general relativistic non-ideal fluid. Finally, we present a simple application of this truncation in the form of a Friedmann-Lema\^{i}tre-Robertson-Walker (FLRW) cosmology.
\par
Throughout this paper, except when otherwise stated, we work in natural units where $c = \hbar = k_\mathrm{B} = 1$.

\section{Relativistic kinetic theory}\label{sec:RelativisticKineticTheory}
\subsection{The relativistic Vlasov equation}
Kinetic theory for classical point particles can be formulated in general coordinates $x$ and in curved space-time.\footnote{For simplicity we assume the absence of the corresponding anti-particles.} In the absence of any force except for gravity, particles follow geodesics,
\begin{equation}
\frac{\differential{^2 x^\mu}}{\differential{s^2}} + \mathit{\Gamma}_{\rho \sigma}^\mu \, \frac{\differential{x^\rho}}{\differential{s}} \frac{\differential{x^\sigma}}{\differential{s}} = 0 \; ,
\end{equation}
where $s$ is an affine parameter and $\mathit{\Gamma}_{\rho \sigma}^\mu(x)$ are the Christoffel symbols of second kind of the space-time metric tensor $g_{\mu \nu}(x)$.\footnote{We work with the metric signature $(-, +, +, +)$.} We denote the components of four-vectors by Greek letters and summation over the same co- and contravariant indices is understood.  The state of the theory is described by a one-particle phase-space distribution function $f(x, p)$. Here, the four-momentum for particles of mass $m$ is ${p^\mu = m \differential{x^\mu} \!/ \! \differential{s}}$. 
\par
The number of particles at position $x$, in the momentum range $\sqrt{g} \, \differential{^4 p} / (2 \pi)^4$ and in a mass range $\differential{m}$ flowing through the hypersurface element
\begin{equation}
\differential{\mathit{\Sigma}_\mu (x)} = \frac{\sqrt{g(x)}}{3!} \, \epsilon_{\mu \nu \rho \sigma} \differential{x^\nu} \differential{x^\rho} \differential{x^\sigma} \, ,
\end{equation}
is given by
\begin{equation}
\begin{multlined}[c][0.88\linewidth]
\differential{N} = \xi(m) \differential{m} p^\mu \differential{\mathit{\Sigma}_\mu(x)} \sqrt{g(x)} \, \frac{\differential{^4 p}}{(2 \pi)^4} \\
\times 4 \pi \theta(p^0) \, \delta \big( p^2 + m^2 \big) \, f(x, p) \; .
\end{multlined}
\end{equation}
Here ${g(x) = -\det(g_{\mu\nu}(x))}$ is the determinant of the metric tensor, $\epsilon_{\mu \nu \rho \sigma}$ is the total antisymmetric Levi-Civita symbol and we abbreviate the space-time inner product as ${p^2 = g_{\mu \nu} p^\mu p^\nu}$. Further $\delta$ and $\theta$ denote the (one-dimensional) Dirac delta and Heaviside unit step function, respectively. For fixed mass $m$, the momenta fulfil the on-shell constraint ${p^2 + m^2 = 0}$, but we allow more generally a distribution of masses $\xi(m)$ which we take to be independent of time and space. As an example for a dark matter model with some distribution of masses one may think about primordial black holes. For a single species of particles with unique mass $m_*$ one has ${\xi(m) = \delta(m - m_*)}$. 
\par
The particle number current is given by
\begin{equation}\label{eq:ParticleNumberCurrent}
N^\mu(x) = \int_m \xi(m) \int_p p^\mu \, 4 \pi \theta(p^0) \, \delta( p^2 + m^2 ) \, f(x, p) \; ,
\end{equation}
and the energy-momentum tensor is
\begin{equation}\label{eq:EnergyMomentumTensor}
T^{\mu \nu}(x) = \!\! \int_m \!\! \xi(m) \! \int_p p^\mu p^\nu \, 4 \pi \theta(p^0) \, \delta( p^2 + m^2 ) \, f(x, p) \; .
\end{equation}
Here and throughout we abbreviate the mass and momentum integrals as
\begin{equation}
\int_m = \int_0^\infty \differential{m} \; , \qquad \int_p = \sqrt{g} \int_{\mathbb{R}^4} \frac{\differential{^4 p}}{(2\pi)^4} \; .
\end{equation}
In the absence of scattering, the distribution function is conserved along a geodesic, ${\differential{f} \! / \! \differential{s} = 0}$, which implies the Vlasov equation \cite{DeGroot1980, Cercignani2002},
\begin{equation}\label{eq:VlasovEq}
\left[ p^\mu \, \frac{\partial}{\partial x^\mu} - \mathit{\Gamma}_{\rho \sigma}^\mu \, p^\rho p^\sigma \, \frac{\partial}{\partial p^\mu} \right] f = 0 \; .
\end{equation}
Note however that this equation is somewhat more general than usual because $f(x,p)$ is here not restricted to ${p^2 + m^2 = 0}$ being satisfied for a unique value of $m$. We allow a more general distribution of masses such that also the distribution function $f(x,p)$ is more general, but equation \eqref{eq:VlasovEq} still determines its time evolution. 
\par
In the following it is convenient to introduce the modified distribution function
\begin{equation}
\tilde{f}(x, p) = \int_m \xi(m) \, 4 \pi \theta(p^0) \, \delta ( p^2 + m^2 ) \, f(x,p) \; ,
\end{equation}
such that for example ${T^{\mu\nu} = \int_p p^\mu p^\nu \tilde{f}(x,p)}$. Equation \eqref{eq:VlasovEq} immediately implies that the modified distribution function also obeys the Vlasov equation,
\begin{equation}
\left[ p^\mu \, \frac{\partial}{\partial x^\mu} - \mathit{\Gamma}_{\rho \sigma}^\mu \, p^\rho p^\sigma \, \frac{\partial}{\partial p^\mu} \right] \tilde{f} = 0 \; .
\end{equation}

\subsection{The method of moments and cumulants}
Often one is not interested in the full distribution function $f(x,p)$ but rather in its moments and cumulants. In analogy to probability distributions one can define fully symmetric four-momentum moments,
\begin{widetext}
\begin{equation}\label{eq:DefMoments}
M^{\mu_1 ... \mu_n}(x) = \int_m \xi(m) \int_p p^{\mu_1} ... p^{\mu_n} \, 4 \pi \theta(p^0) \, \delta ( p^2 + m^2 ) \, f(x, p) = \int_p p^{\mu_1} ... p^{\mu_n} \, \tilde{f}(x, p) \; ,
\end{equation}
the set of which completely characterise the distribution function. These can be conveniently derived from a moment-generating function,
\begin{equation}
z(x; l) = \int_m \xi(m) \int_p \ee^{l_\mu p^\mu} 4 \pi \theta(p^0) \, \delta ( p^2 + m^2 ) \, f(x, p) = \int_p \ee^{l_\mu p^\mu} \tilde{f}(x, p) \; ,
\end{equation}
by $n$-fold differentiation with respect to the source four-vector $l_\mu$. The normalisation ${z(x) = z(x; 0)}$ is the zeroth moment and it is evident from equation \eqref{eq:ParticleNumberCurrent} and \eqref{eq:EnergyMomentumTensor} that the particle number current $N^\mu(x)$ and the energy-momentum tensor $T^{\mu \nu}(x)$ are the first and second moment, respectively. Due to the on-shell Dirac delta function in equation \eqref{eq:DefMoments} one obtains the relation
\begin{equation}\label{eq:OnShellConstraints}
g_{\rho \sigma} M^{\rho \sigma \mu_1 ... \mu_n} + \int_m \xi(m) \, m^2 \int_p p^{\mu_1} ... p^{\mu_n} \, 4 \pi \theta(p^0) \, \delta ( p^2 + m^2 ) \, f(x, p) = 0 \; ,
\end{equation}
\end{widetext}
which in the limit of a single particle species with unique mass $m$ reads
\begin{equation}\label{eq:OnShellConstraintsSingleSpecies}
g_{\rho \sigma} M^{\rho \sigma \mu_1 ... \mu_n} + m^2 \, M^{\mu_1 ... \mu_n} = 0 \; ,
\end{equation}
and relates moments which differ by two orders. For a spectrum of masses it is in general not possible to express the relation \eqref{eq:OnShellConstraints} uniquely in terms of finite moments.
\par
Similarly the distribution function is completely characterised by the connected parts of the moments, the so-called cumulants $C^{\mu_1 ... \mu_n}(x)$. These can be derived from the cumulant-generating function $\ln(z(x; l))$ in the same manner as moments are derived from $z(x; l)$.
\par
The first few cumulants and moments are related by the expressions
\begin{equation}\label{eq:MomentsInCumulants}
\begin{aligned}
N^\mu &= z \, C^\mu \; , \\
T^{\mu \nu} &= z \, (C^{\mu \nu} + C^\mu C^\nu) \; , \\
M^{\mu \nu \rho} &= z \, (C^{\mu \nu \rho} + 3 C^{(\mu \nu} C^{\rho)} + C^\mu C^\nu C^\rho) \; ,
\end{aligned}
\end{equation}
and \textit{vice versa}
\begin{equation}\label{eq:CumulantsInMoments}
\begin{aligned}
C^\mu &= \frac{1}{z} \, N^\mu \; , \\
C^{\mu \nu} &= \frac{1}{z} \, T^{\mu \nu} - \frac{1}{z^2} \, N^\mu N^\nu \; , \\
C^{\mu \nu \rho} &= \frac{1}{z} \, M^{\mu \nu \rho} - \frac{3}{z^2} \, T^{(\mu \nu} N^{\rho)} + \frac{2}{z^3} \, N^\mu N^\nu N^\rho \; .
\end{aligned}
\end{equation}
Here we denote the symmetrisation with respect to a set of indices by a pair of parentheses around them. Similar relations hold for higher order moments and cumulants and can be straight forwardly derived from the corresponding generating functions. Combining the relations \eqref{eq:MomentsInCumulants} and \eqref{eq:CumulantsInMoments} one can express the $n$-th moment in terms of the lower order moments and the $n$-th cumulant,
\begin{equation}\label{eq:MomentsInMoments}
\begin{aligned}
N^\mu &= z \, C^\mu \; , \\
T^{\mu \nu} &= \frac{1}{z} \, N^\mu N^\nu + z \, C^{\mu \nu}, \\
M^{\mu \nu \rho} &= \frac{3}{z} \, T^{(\mu \nu} N^{\rho)}  - \frac{2}{z^2} \, N^\mu N^\nu N^\rho + z \, C^{\mu \nu \rho} \; .
\end{aligned}
\end{equation}
The Vlasov equation \eqref{eq:VlasovEq} implies the covariant conservation of all moments,
\begin{equation}\label{eq:VlasovEqMoments}
\nabla_{\!\nu} M^{\nu \mu_1 ... \mu_n} = 0 \; ,
\end{equation}
where $\nabla_{\!\mu}$ denotes the covariant derivative with respect to the coordinates $x^\mu$. It is immediately clear that the evolution equations are independent of each other at each order.\footnote{However, note that the on-shell constrains \eqref{eq:OnShellConstraints} yield additional relations.} In contrast, the cumulants follow the non-linear evolution equation
\begin{equation}\label{eq:VlasovEqCumulants}
\nabla_{\!\nu} C^{\nu \mu_1 ... \mu_n} + \sum_S C^{\alpha_1 ... \alpha_{\vert S \vert} \nu} \nabla_{\!\nu} C^{\alpha_{{\vert S \vert} + 1} ... \alpha_n} = 0 \; ,
\end{equation}
where the sums runs over all combinations of picking indices $\{ \alpha_1, ..., \alpha_n \}$ out of $\{ \mu_1, ..., \mu_n \}$.\footnote{The sum runs over all $2^n$ sets $S$ in the power set $\mathcal{P}(\{ \mu_1, ..., \mu_n \})$ with indices ${(\alpha_1, ..., \alpha_{\vert S \vert}) \in S}$ and ${(\alpha_{{\vert S \vert} + 1}, ..., \alpha_n) \in \{ \mu_1, ..., \mu_n \} \backslash S}$.} The non-linear terms in the evolution equation of the $n$-th cumulant involve all lower order cumulants and the sum of orders of the two cumulants in the quadratic terms always adds to $n$. This structure sets strong restrictions for a consistent truncation of the cumulant expansion as we discuss at the end of the next section.
\par
We remark that even though the evolution equation \eqref{eq:VlasovEqCumulants} involves all cumulants of lower order it does not couple to higher orders. This is to be seen in contrast to the non-relativistic limit, in which the evolution equations of the moments and cumulants depend on the next higher order moment or cumulant, creating the so-called Vlasov hierarchy \cite{Pueblas2009, Uhlemann2018, Erschfeld2019}. 
\par
This can be made explicit by taking the non-relativistic limit of the moments \eqref{eq:DefMoments}, which for simplicity we do for a single particle species of unique mass $m$. To do so we restore the speed of light $c$ and evaluate the $p^0$ integral to arrive at
\begin{equation}
M^{\mu_1 ... \mu_n} = \sqrt{g} \int_{\mathbb{R}^3} \frac{\differential{^3 p}}{(2 \pi)^3} \frac{p^{\mu_1} ... p^{\mu_n}}{\vert p_0 \vert / c} \, f \; \bigg\vert_{p^0 = E_p / c} \; ,
\end{equation}
where
\begin{equation}
\frac{E_p}{c} = - \frac{g_{0i} p^i}{g_{00}} + \sqrt{\left( \frac{g_{0i} p^i}{g_{00}} \right)^2 - \frac{p_i p^i + m^2 c^2}{g_{00}}} \; ,
\end{equation}
and Latin indices only range over spatial components. In the limit ${c \to \infty}$, where the metric is Minkowskian, the $n$-th moment involving $k$ temporal indices and $n - k$ spatial indices is
\begin{equation}
\frac{1}{c^k} \, M^{0 ... 0 i_1 ... i_{n - k}}(x) \to m^{k - 1} \Mnr^{i_1 ... i_{n - k}}(t, \bm{x}) \; .
\end{equation}
Here $\Mnr^{i_1 ... i_{n - k}}(t, \bm{x})$ is the $(n-k)$-th non-relativistic moment and $t$ denotes time and $\bm{x}$ position in configuration space. From the non-relativistic limit of evolution equation \eqref{eq:VlasovEqMoments} it is then evident, that the evolution of the $n$-th (non-relativistic) moment also depends on the next higher (non-relativistic) moment. A similar argument holds true for the evolution of the non-relativistic cumulants.

\subsection{Truncated cumulant expansion}
In general, the distribution function $f(x,p)$ has an infinite amount of moments, as well as cumulants, the complete set of which encode the same information. Therefore a characterisation in terms of moments or cumulants are two sides of the same coin and are a matter of taste or problem at hand. However, since cumulants of different orders are statistically independent of each other it seems natural to study the cumulant expansion of the distribution function in order to put forth a truncation scheme which allows to describe a finite set of independent degrees of freedom. Assuming one could truncate the expansion after the $n$-th cumulant one would be left with
\begin{equation}\label{eq:IndependentDOF}
\sum_{k=0}^n \begin{pmatrix} 4 + k - 1 \\ k \end{pmatrix} - \sum_{k=1}^n \begin{pmatrix} 4 + k - 2 \\ k - 1 \end{pmatrix} = \begin{pmatrix} 4 + n - 1 \\ n \end{pmatrix}
\end{equation}
independent degrees of freedom. Here, the first sum represents the degrees of freedom from the (fully symmetric) cumulants up to order $n$. However, these are not all independent of each other due to the on-shell constraints \eqref{eq:OnShellConstraints} which reduce the independent degrees of freedom. This is represented by the second sum. 
\par
From the full set of cumulants one can in principle reconstruct the distribution function $\tilde{f}(x,p)$. However, this is not so for a finite truncation, with the exceptions of a truncation after the first or second cumulant, corresponding to a degenerate or normal distribution, respectively.
\par
To work out the consequences of such truncations in more detail it is useful to decompose the particle number current as 
\begin{equation}\label{eq:ParticleNumberCurrentDecomp}
N^\mu = n u^\mu + \nu^\mu \; ,
\end{equation}
and the energy-momentum tensor as
\begin{equation}\label{eq:EnergyMomentumTensorDecomp}
T^{\mu \nu} = \epsilon u^\mu u^\nu + (p + \pib) \Delta^{\mu \nu} + \pi^{\mu \nu} + 2 q^{(\mu} u^{\nu)} \; .
\end{equation}
Here $n$ is the particle number density, $u^\mu$ is the local fluid four-velocity normalised to ${u_\mu u^\mu = -1}$ and $\nu^\mu$ is the diffusion current orthogonal to the fluid velocity, ${u_\mu \nu^\mu = 0}$. Further, $\epsilon$ is the energy density in the local rest frame, $p$ the thermodynamic pressure which is related to $n$ and $\epsilon$ by the equilibrium expression, $\pib$ is the bulk viscous pressure, ${\Delta^{\mu \nu} = u^\mu u^\nu + g^{\mu \nu}}$ is a projector orthogonal to the fluid velocity, $\pi^{\mu \nu}$ is the shear stress tensor, which is symmetric, transverse to the fluid velocity $u_\mu \pi^{\mu \nu} = 0$ and traceless ${\tensor{\pi}{_\mu^\mu} = 0}$ and $q^\mu$ is the heat current which is orthogonal to the fluid velocity, ${u_\mu q^\mu = 0}$. In the following we abbreviate the sum of thermodynamic and bulk viscous pressure as an effective pressure, ${\peff = p + \pib}$. These rather general relations can be specialised to a frame by fixing the fluid four-velocity. Common choices are the Landau frame \cite{Landau1959}, where the fluid four-velocity is a time-like eigenvector of the energy-momentum tensor and the heat current vanishes, or the Eckart frame \cite{Eckart1940}, in which the fluid four-velocity is defined by the direction of the particle number current so that one has vanishing diffusion current.
\par
In the most simple non-trivial case one truncates the cumulant expansion after the first order and obtains the so-called single-stream approximation. It is characterised by four independent degrees of freedom and for a single particle species with unique mass $m$ the on-shell constraint \eqref{eq:OnShellConstraintsSingleSpecies} leads to ${z = n / m}$. The modified distribution function is a degenerate distribution,
\begin{equation}\label{eq:DegenerateTruncation}
\tilde{f} = \frac{(2 \pi)^4}{\sqrt{g}} \frac{n}{m} \, \delta^{(4)}(p - m u) \; ,
\end{equation}
where $\delta^{(4)}$ denotes the four-dimensional Dirac delta function. In this case one has ${\epsilon = n m}$ and $\peff = \nu^\mu = q^\mu = \pi^{\mu\nu} = 0$, or in other words, the particle number current and the energy-momentum tensor are the ones of an perfect pressureless fluid.
\par
Including the second cumulant and truncating the expansion at third order one obtains a normal distribution characterised by ten independent degrees of freedom.\footnote{A general (unnormalised) four-dimensional normal distribution is as usual characterised by cumulants up to second order, corresponding to 15 degrees of freedom. However, the on-shell constraints \eqref{eq:OnShellConstraints} reduce the independent degrees of freedom to ten, see also equation \eqref{eq:IndependentDOF} and the discussion thereafter.} Again specialising to a single particle species with unique mass $m$, the second moment on-shell constraint ${g_{\mu\nu} T^{\mu\nu} = m^2 z}$ leads to
\begin{equation}\label{eq:NormalTruncationOnShell1}
z = \frac{\epsilon - 3 \peff}{m^2} \; .
\end{equation}
The third moment on-shell constraint projected along the fluid velocity ${u_\rho g_{\mu\nu} M^{\mu\nu\rho} = m^2 u_\rho N^\rho}$ gives
\begin{equation}\label{eq:NormalTruncationOnShell2}
\frac{2}{z^3} \, [ n^3 - z n \epsilon - n \nu_\mu \nu^\mu + z q_\mu \nu^\mu ] = g_{\mu \nu} C^{\mu \nu \rho} u_\rho = 0 \; ,
\end{equation}
while the projection orthogonal to the fluid velocity ${\tensor{\Delta}{^\mu_\sigma} g_{\nu\rho} M^{\nu\rho\sigma} = m^2 \tensor{\Delta}{^\mu_\sigma} N^\sigma}$ yields
\begin{equation}\label{eq:NormalTruncationOnShell3}
\begin{multlined}[c][0.88\linewidth]
\frac{2}{z^3} \, [ (\nu_\nu \nu^\nu - n^2 - z \peff) \nu^\mu - z \pi^{\mu \nu} \nu_\nu + z n q^\mu ] \\
= \tensor{\Delta}{^\mu_\sigma} g_{\nu \rho} C^{\nu \rho \sigma} = 0 \; .
\end{multlined}
\end{equation}
The modified distribution function is given by,
\begin{equation}\label{eq:NormalTruncation}
\begin{multlined}[c][0.88\linewidth]
\tilde{f} = \frac{(2 \pi)^4}{\sqrt{g}} \frac{n^2}{\epsilon} \, \delta \left( u \cdot p - \frac{\epsilon}{n} \right) \frac{1}{(2 \pi)^{\frac{3}{2}} \det(C_{\alpha\beta})^{\frac{1}{2}}} \\
\times \exp\left\{ - \frac{1}{2} \, p^\mu \, (C^{-1})_{\mu \nu} \, p^\nu \right\} \; ,
\end{multlined}
\end{equation}
where ${u \cdot p = u_\mu p^\mu}$ and due to the constraint \eqref{eq:NormalTruncationOnShell3} the second cumulant is purely transverse to the fluid velocity,
\begin{equation}
C^{\mu \nu} = \peff \, \Delta^{\mu \nu} + \pi^{\mu \nu} \; ,
\end{equation}
and therefore the parallel part collapses to a Dirac delta function. The constraints \eqref{eq:NormalTruncationOnShell1} and \eqref{eq:NormalTruncationOnShell2} can be combined to give
\begin{equation}
\frac{n^2}{\epsilon} = \frac{\epsilon - 3 \peff}{m^2} \; ,
\end{equation}
reducing the independent degrees of freedom to ten. Interestingly enough this truncation implies vanishing diffusion and heat current and thus the Landau and Eckart frame are the same. Formally we have now a fluid with non-vanishing bulk viscous pressure and shear stress.
\par
This truncation scheme can be straight forwardly generalised to higher orders. However, there are no distributions which are characterised by a finite number of cumulants beside the degenerate and normal distributions and one is therefore not able to explicitly reconstruct $\tilde{f}(x,p)$. Further one cannot find a corresponding distribution function $f(x,p)$ because one is not able reconstruct the $\delta(p^2 + m^2)$. Nevertheless, as an approximate model for a more complex form, equation \eqref{eq:NormalTruncation} may be quite reasonable. A nice feature is that the cumulants that govern it have rather transparent equations of motion that can be solved in the spirit of a non-ideal fluid approximation as we discuss in the next section.
\par
While we are in principle able to truncate the cumulant expansion of the distribution function, the questions arises whether these truncations are preserved by the Vlasov equation \eqref{eq:VlasovEqCumulants}. Interestingly enough the single-stream approximation \eqref{eq:DegenerateTruncation} is (apparently) preserved since no higher order cumulants are sourced by terms solely depending on the zeroth or first cumulant. However, this apparent self-consistency is not stable under perturbations, that is as soon as any other cumulant obtains a non-vanishing value, cumulants of all orders are generated. This can be checked from the combinations of cumulants appearing in the quadratic terms of equation \eqref{eq:VlasovEqCumulants}. This also implies that any truncation beyond the single-stream approximation is not preserved by the Vlasov equation and higher order cumulants are generated throughout evolution in time. The self-consistency of the single-stream approximation is only apparent due to the phenomenon of shell-crossing, when multiple streams of matter coexist at the same region in space. At this point in configuration space the velocity field in equation \eqref{eq:DegenerateTruncation} is multivalued and the corresponding distribution function $\tilde{f}(x,p)$ has non-vanishing second and higher order cumulants \cite{Pueblas2009}.

\section{Closed system of equations}\label{sec:ClosedSystemOfEq}
\subsection{Evolution equations}
The particle number density and energy-momentum tensor have 14 independent degrees of freedom which we parametrise in terms of the fields introduced in equation \eqref{eq:ParticleNumberCurrentDecomp} and \eqref{eq:EnergyMomentumTensorDecomp}. Evolution equations for these can be obtained from the covariant conservation of the first three moments,
\begin{equation}
\nabla_{\!\mu} N^\mu = 0 \; , \qquad \nabla_{\!\nu} T^{\mu \nu} = 0 \; , \qquad \nabla_{\!\rho} M^{\mu \nu \rho} = 0 \; ,
\end{equation}
by expressing the third moment in terms of lower order moments and the third cumulant, as done in equation \eqref{eq:MomentsInMoments}.
\par
The conservation of the particle number current yields an evolution equation for the particle number density,
\begin{equation}\label{eq:ParticleNumberDensityEvolution}
u^\mu \nabla_{\!\mu} n + n \nabla_{\!\mu} u^\mu + \nabla_{\!\mu} \nu^\mu = 0 \; .
\end{equation}
Projecting the conservation of the energy-momentum tensor along the fluid four-velocity, ${u_\mu \nabla_{\!\nu} T^{\mu \nu} = 0}$, gives an evolution equation for the energy density,
\begin{widetext}
\begin{equation}\label{eq:EnergyDensityEvolution}
u^\mu \nabla_{\!\mu} \epsilon + (\epsilon + \peff) \nabla_{\!\mu} u^\mu + \pi^{\mu\nu} \nabla_{\!\mu} u_\nu + \nabla_{\!\mu} q^\mu + q^\nu u^\mu \nabla_{\!\mu} u_\nu = 0 \; ,
\end{equation}
while a projection orthogonal, ${\tensor{\Delta}{^\mu_\nu} \nabla_{\!\rho} T^{\nu \rho} = 0}$, yields essentially an evolution equation of the fluid four-velocity,
\begin{equation}\label{eq:FluidVelocityEvolution}
(\epsilon + \peff) u^\nu \nabla_{\!\nu} u^\mu + \Delta^{\mu\nu} \nabla_{\!\nu} \peff + \tensor{\Delta}{^\mu_\nu} \nabla_{\!\rho} \pi^{\rho\nu} + \tensor{\Delta}{^\mu_\nu} u^\rho \nabla_{\!\rho} q^\nu + q^\mu \nabla_{\!\nu} u^\nu + q^\nu \nabla_{\!\nu} u^\mu = 0 \; .
\end{equation}
A projection onto the orthogonal parts of the covariant conservation of the third moment, ${\Delta_{\mu\nu} \nabla_{\!\rho} M^{\mu\nu\rho} = 0}$, gives an evolution equation for the effective pressure,
\begin{equation}\label{eq:EffectivePressureEvolution}
\begin{multlined}[c][0.88\linewidth]
(n u^\rho + \nu^\rho) \left\{ 3 \nabla_{\!\rho} \peff + 2 q_\mu \nabla_{\!\rho} u^\mu + (4 \nu_\mu \nu^\mu - 3 z \peff) \, \frac{\nabla_{\!\rho} z}{z^2} - \frac{4}{z} \, (n \nu_\mu \nabla_{\!\rho} u^\mu + \nu_\mu \nabla_{\!\rho} \nu^\mu) \right\} \\
- 2 (\peff \nu^\rho + \nu_\mu \pi^{\mu\rho} + q_\mu \nu^\mu u^\rho) \, \frac{\nabla_{\!\rho} z}{z} + 2 (\peff \Delta^{\mu\rho} + \pi^{\mu\rho} + q^\mu u^\rho) \bigg\{ n \nabla_{\!\rho} u_\mu + \nabla_{\!\rho} \nu_\mu \bigg\} \\[0.5ex]
= - z \Delta_{\mu\nu} \Big\{ z \nabla_{\!\rho} C^{\mu \nu \rho} + C^{\mu \nu \rho} \nabla_{\!\rho} z \Big\} \; .
\end{multlined}
\end{equation}
The evolution equation for the three degrees of freedom parametrised by the diffusion current $\nu^\mu$ and heat current $q^\mu$ is obtained from the projection ${\tensor{\Delta}{^\alpha_\mu} u_{\nu}\nabla_{\!\rho} M^{\mu\nu\rho} = 0}$,
\begin{equation}\label{eq:DiffusionHeatCurrentEvolution}
\begin{aligned}
&(n u^\rho + \nu^\rho) \bigg\{ \tensor{\Delta}{^\alpha_\mu} \nabla_{\!\rho} q^\mu + (\epsilon + \peff) \nabla_{\!\rho} u^\alpha + \pi^{\alpha\mu} \nabla_{\!\rho} u_\nu + (4 n \nu^\alpha - z q^\alpha) \, \frac{\nabla_{\!\rho} z}{z^2} \\
&\hspace*{17.2em}- \frac{2}{z} \, (n^2 \nabla_{\!\rho} u^\alpha + \tensor{\Delta}{^\alpha_\mu} n \nabla_{\!\rho} \nu^\mu + \nu^\alpha \nabla_{\!\rho} n + \nu^\alpha \nu^\nu \nabla_{\!\rho} u_\nu) \bigg\} \\[0.5ex]
+ \;&(\epsilon u^\rho + q^\rho) \left\{ \tensor{\Delta}{^\alpha_\mu} \nabla_{\!\rho} \nu^\mu + n \nabla_{\!\rho} u^\alpha - \nu^\alpha \, \frac{\nabla_{\!\rho} z}{z} \right\} + (\peff \Delta^{\rho\alpha} + \pi^{\rho\alpha} + u^\rho q^\alpha) \left\{ \nabla_{\!\rho} n + \nu^\nu \nabla_{\!\rho} u_\nu - n \, \frac{\nabla_{\!\rho} z}{z} \right\} \\[0.5ex]
&\hspace*{30em}= z \tensor{\Delta}{^\alpha_\mu} u_\nu \Big\{ z \nabla_{\!\rho} C^{\mu\nu\rho} + C^{\mu\nu\rho} \nabla_{\!\rho} z \Big\} \; .
\end{aligned}
\end{equation}
Finally, the evolution for the shear stress tensor $\pi^{\mu\nu}$ follows from the projection ${\tensor{P}{^{\alpha\beta}_{\mu\nu}} \nabla_{\!\rho} M^{\mu\nu\rho} = 0}$,
\begin{equation}\label{eq:ShearStressEvolution}
\begin{multlined}[c][0.88\linewidth]
(n u^\rho + \nu^\rho) \tensor{P}{^{\alpha\beta}_{\mu\nu}} \left\{ \nabla_{\!\rho} \pi^{\mu\nu} + 2 q^\nu \nabla_{\!\rho} u^\mu + (4 \nu^\mu \nu^\nu - z \pi^{\mu\nu}) \, \frac{\nabla_{\!\rho} z}{z^2} \right\} \\
+ 2 (\peff \Delta^{\mu\rho} + \pi^{\mu\rho} + q^\mu u^\rho) \tensor{P}{^{\alpha\beta}_{\mu\nu}} \left\{ n \nabla_{\!\rho} u^\nu + \nabla_{\!\rho} \nu^\nu - \nu^\nu \, \frac{\nabla_{\!\rho} z}{z} \right\} \\
= - z \tensor{P}{^{\alpha\beta}_{\mu\nu}} \Big\{ z \nabla_{\!\rho} C^{\mu\nu\rho} + C^{\mu\nu\rho} \nabla_{\!\rho} z \Big\} \; ,
\end{multlined}
\end{equation}
\end{widetext}
where the symmetric, traceless and transverse projector reads
\begin{equation}
\tensor{P}{^{\alpha\beta}_{\mu\nu}} = \tfrac{1}{2} \, \tensor{\Delta}{^{\alpha}_{\mu}} \tensor{\Delta}{^{\beta}_{\nu}} + \tfrac{1}{2} \, \tensor{\Delta}{^{\alpha}_{\nu}} \tensor{\Delta}{^{\beta}_{\mu}} - \tfrac{1}{3} \, \tensor{\Delta}{^{\alpha\beta}_{}} \tensor{\Delta}{^{}_{\mu\nu}} \; .
\end{equation}
It is evident that the system of equations \eqref{eq:ParticleNumberDensityEvolution} -- \eqref{eq:ShearStressEvolution} is not closed, since they couple to the normalisation $z$ and the third cumulant $C^{\mu\nu\rho}$. In order to close the system of equations we impose a truncation of the cumulant expansion after the second cumulant, $C^{\mu\nu\rho} = 0$, and use the energy-momentum tensor on-shell constraint \eqref{eq:NormalTruncationOnShell1} to eliminate $z$. Strictly speaking not all of the equations are independent of each other due to equations \eqref{eq:NormalTruncationOnShell2} and \eqref{eq:NormalTruncationOnShell3}. These reduce the independent degrees of freedom to ten in accordance with formula \eqref{eq:IndependentDOF}. Nevertheless, we assume that these constraints can be neglected with the reasoning as follows. We know from the discussion at the end of section \ref{sec:RelativisticKineticTheory} that the imposed truncation is not preserved by the Vlasov equation and the third cumulant is generated throughout the evolution in time. Therefore the constraints \eqref{eq:NormalTruncationOnShell2} and \eqref{eq:NormalTruncationOnShell3} are not strictly satisfied but rather proportional to the third cumulant as indicated on the right-hand side. That is, we do not assume the third cumulant to be exactly zero throughout time evolution, but rather to stay small enough to be neglected in the evolution equations.
\par
A similar truncation of the cumulant expansion after the second order has also been used in the non-relativistic limit to model dark matter with non-vanishing velocity dispersion \cite{McDonald2011, Erschfeld2019}.
\par
Collecting the 14 degrees of freedom in a superfield $\mathit{\Phi}^a(x)$ the evolution equations \eqref{eq:ParticleNumberDensityEvolution} -- \eqref{eq:ShearStressEvolution} can be written in the quasi-linear form
\begin{equation}
\mathcal{A}_a u^\rho \nabla_{\!\rho} \mathit{\Phi}^a + \tensor{\mathcal{B}}{^\rho_a} \nabla_{\!\rho} \mathit{\Phi}^a = 0 \; .
\end{equation}
Here $\mathcal{A}_a$ and $\tensor{\mathcal{B}}{^\rho_a}$ are field dependent matrices and ${u_\rho \tensor{\mathcal{B}}{^\rho_a} = 0}$. The index $a$ carries the appropriate amount of covariant indices, e.g. in the Eckart frame one has ${\mathit{\Phi}^a = (n, \epsilon, \peff, u^\mu, q^\mu, \pi^{\mu\nu})}$. Since the expressions for the matrices $\mathcal{A}_a$ and $\tensor{\mathcal{B}}{^\rho_a}$ are less transparent and rather cumbersome if written explicit, we restrain from displaying them. We checked that the matrix $\mathcal{A}_a$ is diagonalisable and invertible, making the system hyperbolic \cite{Floerchinger2018}. Finally, we are left with a closed, covariant and hyperbolic system of equations which describe the 14 degrees of freedom introduced in equations \eqref{eq:ParticleNumberCurrentDecomp} and \eqref{eq:EnergyMomentumTensorDecomp}.

\subsection{FLRW cosmology}
\begin{figure}
\centering
\includegraphics[width=8.6cm]{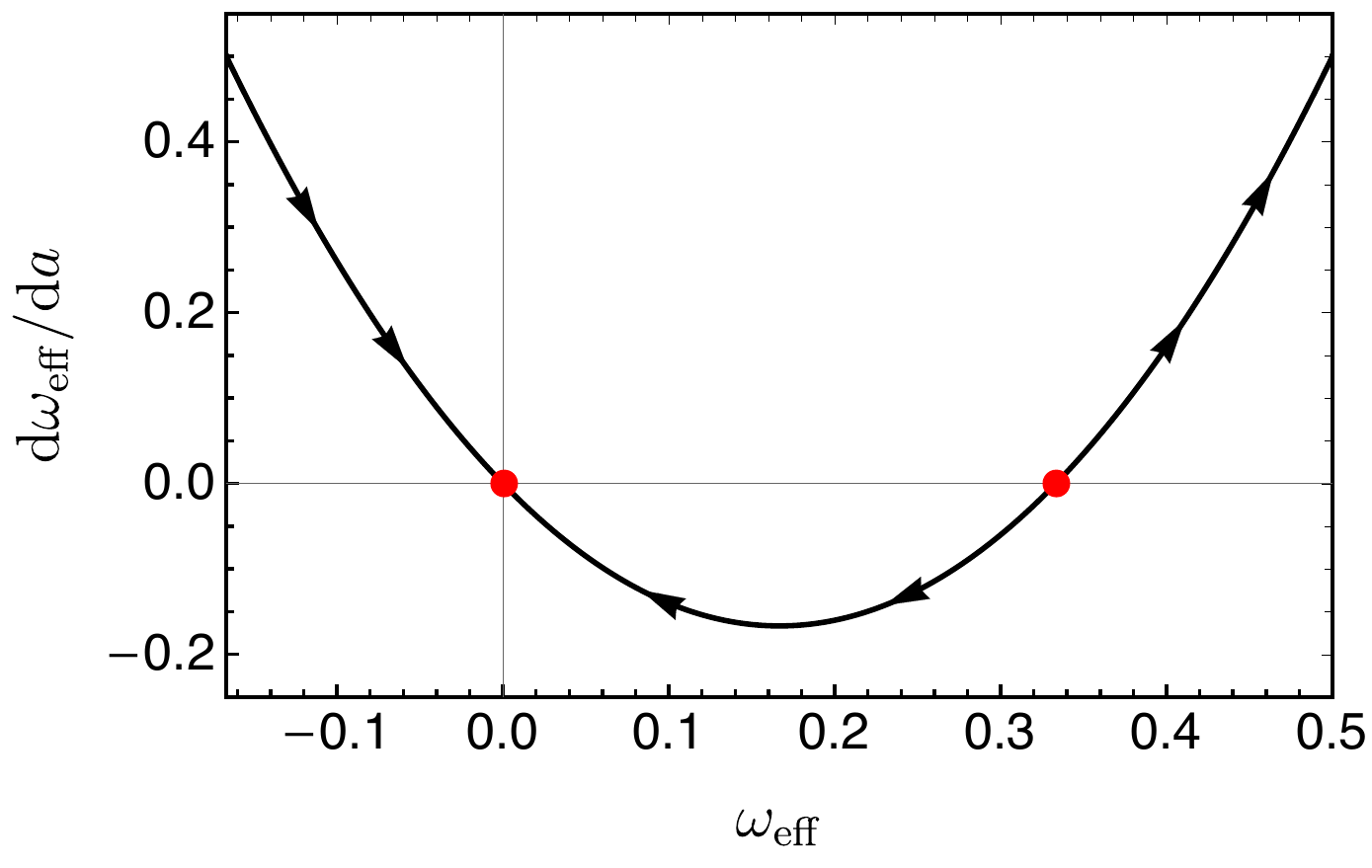}
\caption{The flow of the effective equation of state $\omegaeff$ is indicated by arrows and the fix points as red dots. It has an attractive fixed point at ${\omegaeff = 0}$ corresponding to non-relativistic matter as well as a repulsive fixed point at ${\omegaeff = 1/3}$ corresponding to ultra-relativistic radiation.}
\label{fig:EquationOfState}
\end{figure}

As a simple cosmological working example we consider a spatially flat Friedmann-Lema\^{i}tre-Robertson-Walker metric with line element
\begin{equation}
\differential{s^2} = - \differential{t^2} + a(t)^2 \, \delta_{ij} \differential{x^i} \differential{x^j} \, ,
\end{equation}
where $a(t)$ is the dimensionless scale factor which parametrises the relative spatial expansion of the Universe. Due to the symmetries of the metric, namely homogeneity and isotropy, the particle number current and energy-momentum tensor have the form of a perfect fluid,
\begin{equation}
N^\mu = n u^\mu \; , \qquad T^{\mu \nu} = \epsilon u^\mu u^\nu + \peff \, \Delta^{\mu \nu} \; .
\end{equation}
Here the energy density $\epsilon(t)$ and effective pressure $\peff(t)$ are functions of time only and the four-velocity is given by ${u^\mu = (1, 0, 0, 0)}$. We define the effective equation of state parameter as the ratio of the effective pressure and energy density, ${\omegaeff = \peff / \epsilon}$. The evolution equations \eqref{eq:ParticleNumberDensityEvolution} and \eqref{eq:EnergyDensityEvolution} read
\begin{equation}
\partial_t n + 3 H n = 0 \; , \qquad \partial_t \epsilon + 3 H (1 + \omegaeff) \epsilon = 0 \; ,
\end{equation}
while equation \eqref{eq:EffectivePressureEvolution} gives
\begin{equation}
\partial_t \omegaeff + 2 H (1 - 3 \omegaeff) \omegaeff = 0 \; ,
\end{equation}
and ${H = \partial_t a / a}$ is the Hubble rate. These equations can be solved in terms of the scale factor,
\begin{equation}
\begin{aligned}
n &= n_0 \, a^{-3} \; , \\
\epsilon &= \epsilon_0 \, a^{-4} \left[ a^2 (1 - 3 \omegaeffo) + 3 \omegaeffo \right]^{\frac{1}{2}} \; , \\
\omegaeff &= \omegaeffo \left[ a^2 (1 - 3 \omegaeffo) + 3 \omegaeffo \right]^{-1} \; ,
\end{aligned}
\end{equation}
where quantities subscripted with a $0$ are the values at ${a(t) = 1}$, corresponding to today. The particle number density decays as expected with the expansion of space while the decay of the energy density also depends on the effective equation of state parameter. Figure \ref{fig:EquationOfState} displays the growth of the effective equation of state parameter as a function of itself. The arrows indicate the flow of the solution and the red dots indicate the fixed points. There is an attractive fixed point at ${\omegaeff = 0}$ corresponding to non-relativistic matter and a repulsive fixed point at ${\omegaeff = 1/3}$ corresponding to ultra-relativistic radiation. For an effective equation of state parameter ${\omegaeff < 1/3}$ the solution evolves towards the attractive non-relativistic solution fixed point as is expected for matter that is non-interacting except for gravity. For an effective equation of state parameter ${\omegaeff > 1/3}$ the solution exhibits a strong growth, but the physical interpretation behind such values of the equation of state parameter is not clear.
\par
In a next step it would be interesting to treat non-homogeneous solutions and cosmological structure formation. The full set of non-ideal fluid fields, including the peculiar fluid velocity, shear stress and diffusion or heat current, are then expected to become non-trivial. Interestingly, perturbations in these fields can also influence the overall cosmological expansion through a dissipative back-reaction \cite{Floerchinger2015}.

\section{Conclusions}
Starting from a general relativistic kinetic theory approach for a system of collisionless classical point particles we presented the method of moments and cumulants of the one-particle phase-space distribution function. We studied how the moments and cumulants evolve under the relativistic Vlasov equation and found in particular that all moments are covariantly conserved while the cumulants follow a more complex non-linear evolution equation. For the first and second moment these are the common covariant conservation laws related to the particle number current and energy-momentum tensor, but the evolution of higher order moments furnish additional evolution equations.
\par
We showed how the cumulant expansion of the distribution function can be truncated at finite order and explicitly performed this truncation after the first and second cumulant, corresponding to the single-stream and a Gaussian approximation, respectively. In particular the Gaussian approximation is capable of describing a non-ideal fluid with non-vanishing bulk viscous pressure and shear stress. We discussed that these kind of truncations are not preserved by the Vlasov equation since higher order cumulants are naturally generated by lower order cumulants.
\par
From the covariant conservation of the first three moments we derived a closed, covariant and hyperbolic system of equations by neglecting the third cumulant. The equations give the evolution of the 14 degrees of freedom of the particle number current and energy-momentum tensor and can be used to describe a general relativistic non-ideal fluid. As a working example we considered a Friedmann-Lema\^{i}tre-Robertson-Walker cosmology with non-vanishing dynamic pressure and solve its time evolution. We find that the solution has an attractive and repulsive fixed point, corresponding to non-relativistic matter and ultra-relativistic radiation, respectively.
\par
The equations of motion for a non-ideal fluid approximation to dark matter that we have derived can be extended in different directions. One would be to include the effects of dark matter self interaction or interactions with baryonic matter. Another would be to include quantum effects. Moreover, it would of course be highly interesting to solve the evolution equations for the non-ideal fluid fields or cumulants directly, either numerically or with further analytical techniques such as perturbation theory or field theoretic methods similar to those developed in references~\cite{Blas2015, Floerchinger2017, Floerchinger2019}. In particular we are curious to see whether the truncated cumulant expansion developed here, or an extension of it, can agree with numerical solutions of the Vlasov equation through $N$-body simulations. This might lead to a rather useful framework to study extensions of the collisionless cold dark matter model and for a comparison to observational data. 

\begin{acknowledgments}
The authors thank E.~Grossi for useful discussions. This work is supported by the Deutsche Forschungsgemeinschaft (German Research Foundation) under Germany's Excellence Strategy and the Cluster of Excellence EXC~2181 (STRUCTURES), the Collaborative Research Centre SFB~1225 (ISOQUANT) as well as the research grant FL~736/3-1.
\end{acknowledgments}

\appendix
\section{Relaxation-time approximation}
In a setting with collisions, as for example for self-interacting dark matter, the evolution of the distribution function is determined by the relativistic Boltzmann equation \cite{DeGroot1980, Cercignani2002}
\begin{equation}
\left[ p^\mu \, \frac{\partial}{\partial x^\mu} - \mathit{\Gamma}_{\rho \sigma}^\mu \, p^\rho p^\sigma \, \frac{\partial}{\partial p^\mu} \right] f = \mathcal{C}[f] \; ,
\end{equation}
which differs from the Vlasov equation \eqref{eq:VlasovEq} by the collision integral $C[f]$, which in general is a non-linear functional of the distribution function. The situation is greatly simplified in the relaxation-time approximation \cite{Bhatnagar1954, Anderson1974}, for which the collision terms is
\begin{equation}
\mathcal{C} = - (u \cdot p) \, \frac{f - \feq}{\taueq} \; .
\end{equation}
Here $\feq(x,p)$ is the equilibrium distribution function and $\taueq$ the relaxation time. Defining the equilibrium moments $\Meq^{\mu_1 ... \mu_n}(x)$ in the same manner as done in equation \eqref{eq:DefMoments} for the equilibrium distribution function, one can derive the evolution equation
\begin{equation}\label{eq:BoltzmannEqMoments}
\nabla_{\!\nu} M^{\nu \mu_1 ... \mu_n} = - \frac{u_\nu}{\taueq} \left[ M^{\nu \mu_1 ... \mu_n} - \Meq^{\nu \mu_1 ... \mu_n} \right] \; .
\end{equation}
While the moments are no longer covariantly conserved, they are still independent of each other at each order. Since the particle number current and energy-momentum tensor are still covariantly conserved, we can infer their equilibrium form,
\begin{equation}
\Neq^\mu = n u^\mu \; , \qquad \Teq^{\mu \nu} = \epsilon u^\mu u^\nu + p \Delta^{\mu \nu} \; .
\end{equation}
This also fixes the equilibrium normalisation to ${\zeq = (\epsilon - 3p) / m^2}$. Proceeding in the same manner as done in section \ref{sec:ClosedSystemOfEq} to derive equations \eqref{eq:ParticleNumberDensityEvolution} -- \eqref{eq:ShearStressEvolution}, the equations \eqref{eq:ParticleNumberDensityEvolution} -- \eqref{eq:FluidVelocityEvolution} are unchanged due to the covariant conservation of the particle number current and energy-momentum tensor. However, the equations \eqref{eq:EffectivePressureEvolution} -- \eqref{eq:ShearStressEvolution} obtain additional contributions due to the terms on the right-hand side of equation \eqref{eq:BoltzmannEqMoments} stemming from the relaxation-time approximation of the collision integral. The additional terms appearing on the right-hand sides of equations \eqref{eq:EffectivePressureEvolution}, \eqref{eq:DiffusionHeatCurrentEvolution} and \eqref{eq:ShearStressEvolution} are
\begin{equation}
\begin{multlined}[c][0.87\linewidth]
- \frac{1}{\taueq} \bigg[ 3 n \left( \frac{z}{\zeq} \, p - \peff \right) + 2 \left( \frac{n}{z} \, \nu_\mu - q_\mu \right) \nu^\mu \\
+ z \Delta_{\mu\nu} u_\rho (z \, C^{\mu\nu\rho} - \zeq \Ceq^{\mu\nu\rho}) \bigg] \; ,
\end{multlined}
\end{equation}
\begin{equation}
\begin{multlined}[c][0.87\linewidth]
- \frac{1}{\taueq} \bigg[ \left( \frac{2 n}{z} \, n - \epsilon \right) \nu^\alpha - 2 n q^\alpha \\
- z \tensor{\Delta}{^\alpha_\mu} u_\nu u_\rho (z \, C^{\mu\nu\rho} - \zeq \Ceq^{\mu\nu\rho}) \bigg] \; ,
\end{multlined}
\end{equation}
and
\begin{equation}
\begin{multlined}[c][0.87\linewidth]
- \frac{1}{\taueq} \bigg[ 2 \tensor{P}{^{\alpha\beta}_{\mu\nu}} \left( \frac{n}{z} \, \nu^\mu - q^\mu \right) \nu^\nu - n \pi^{\alpha\beta} \\
+ z \tensor{P}{^{\alpha\beta}_{\mu\nu}} u_\rho (z \, C^{\mu\nu\rho} - \zeq \Ceq^{\mu\nu\rho}) \bigg] \; ,
\end{multlined}
\end{equation}
respectively. This approach may be generalised to more complex collision integrals, such as the Sto{\ss}zahlansatz, although it is not clear whether one can explicitly solve the integral in this case. For the Gaussian approximation employed in this work one could hope to have a chance at solving the collision integral, although it heavily depends on the explicit form of the collision kernel.

\end{document}